\documentclass[a4paper,fleqn]{cas-dc}

\usepackage{subcaption} 
\usepackage[numbers,sort&compress]{natbib}
\usepackage{cleveref}
\crefname{figure}{Fig.}{Figs.}
\crefname{equation}{Eq.}{Eqs.}

\usepackage{algorithm}
\usepackage{algorithmic}

\def\tsc#1{\csdef{#1}{\textsc{\lowercase{#1}}\xspace}}
\tsc{WGM}
\tsc{QE}
\tsc{EP}
\tsc{PMS}
\tsc{BEC}
\tsc{DE}

\begin{document}
\let\WriteBookmarks\relax
\def\floatpagepagefraction{1}
\def\textpagefraction{.001}

\shorttitle{Spatial public goods games with queueing and reputation}

\shortauthors{Gui zhang et~al.}

\title [mode = title]{Spatial public goods games with queueing and reputation}                      

%



\credit{Conceptualization of this study, Methodology, Software}

\author[add1]{Gui Zhang}[style=Chinese]
\author[add1]{Xiaojin Xiong}[style=Chinese]
\author[add2]{Bin Pi}[style=Chinese]
\author[add1]{Minyu Feng}[style=Chinese,orcid=0000-0001-6722-3017]
\cormark[1]
\ead{myfeng@swu.edu.cn}

\author[add3,add4,add5,add6,add7]{Matja{\v z} Perc}[orcid=0000-0002-3087-541X]
\cormark[1]
\ead{matjaz.perc@gmail.com}
\affiliation[add1]{organization={College of Artificial Intelligence},
	addressline={Southwest University}, 
	city={Chongqing},
	postcode={400715}, 
	country={China}
    }
\affiliation[add2]{organization={School of Mathematical Sciences, University of Electronic Science and Technology of China}, 
    city={Chengdu}, 
    postcode={611731}, 
    country={China}
}
\affiliation[add3]{organization={Faculty of Natural Sciences and Mathematics, University of
Maribor},
	addressline={Koro{\v s}ka cesta 160}, 
	city={Maribor},
	postcode={2000}, 
	country={Slovenia}
}

\affiliation[add4]{organization={Community Healthcare Center Dr. Adolf Drolc Maribor},
    addressline={Ulica talcev 9}, 
    city={Maribor},
    postcode={2000}, 
    country={Slovenia}
}

\affiliation[add5]{organization={Department of Physics, Kyung Hee University},
    addressline={26 Kyungheedae-ro, Dongdaemun-gu}, 
    postcode={02447}, 
    city={Seoul}, 
    country={Republic of Korea}
}

\affiliation[add6]{organization={Complexity Science Hub},
    addressline={Metternichgasse 8}, 
    postcode={1030}, 
    city={Vienna}, 
    country={Austria}
}

\affiliation[add7]{organization={University College, Korea University},
    addressline={145 Anam-ro, Seongbuk-gu}, 
    postcode={02841}, 
    city={Seoul}, 
    country={Republic of Korea}
}



\begin{abstract}
In real-world social and economic systems, the provisioning of public goods generally entails continuous interactions among individuals, with decisions to cooperate or defect being influenced by dynamic factors such as timing, resource availability, and the duration of engagement. However, the traditional public goods game ignores the asynchrony of the strategy adopted by players in the game. To address this problem, we propose a spatial public goods game that integrates an $M/M/1$ queueing system to simulate the dynamic flow of player interactions. We use a birth-death process to characterize the stochastic dynamics of this queueing system, with players arriving following a Poisson process and service times being exponentially distributed under a first-come-first-served basis with finite queue capacity. We also incorporate reputation so that players who have cooperated in the past are more likely to be chosen for future interactions. Our research shows that a high arrival rate, low service rate, and the reputation mechanism jointly facilitate the emergence of cooperative individuals in the network, which thus provides an interesting and new perspective for the provisioning of public goods.
\end{abstract}




\maketitle
\section{Introduction}

In societies, individual behavior often exerts a significant impact on the dynamics of the group, either promoting or inhibiting societal development. Generally, the interests of the individual may conflict with those of the group, leading to the emergence of a cooperation dilemma \cite{rand2013human,melis2010human}. Fortunately, evolutionary game theory is used to describe the conflict between individuals' choices and population interests \cite{weibull1997evolutionary,sigmund1999evolutionary,perc2010coevolutionary}, which are extensively studied in biology \cite{traulsen2023future,leimar2023game}, economics \cite{kandori1997evolutionary,martcheva2021effects}, and sociology \cite{grigolini2017call}. The prisoner's dilemma game (PDG) \cite{axelrod1980more, yao2023inhibition}, snow drift game (SDG) \cite{hauert2004spatial, pi2022evolutionary}, and hawk-dove game (HDG) \cite{auger1998hawk, heller2024stable} have become representative models for describing social dilemmas, which have stimulated extensive research on evolutionary game dynamics in various network structures, such as square lattice networks \cite{szabo1998evolutionary}, small-world networks \cite{liu2024evolution, deng2010memory}, and scale-free networks. These game models have promoted the study of the evolution of cooperation in social groups \cite{ohtsuki2006simple,nowak2004evolutionary}. 
In order to solve the dilemma in these games, many mechanisms have been widely studied to improve the level of cooperation among the population. For instance, Nowak and May studied a spatial version of the PDG and found that cooperative and defective individuals can coexist for a long time \cite{nowak1992evolutionary}. Furthermore, Nowak summarized kin selection, direct reciprocity, indirect reciprocity, network reciprocity, and group selection \cite{nowak2006five} and derived a simple rule to determine whether natural selection can promote the evolution of cooperation \cite{ohtsuki2006simple}. At the same time, a large number of scholars have designed mechanisms including reward \cite{motepalli2021reward,meng2021dynamic}, reputation \cite{tian2019evaluating}, environmental feedback \cite{ding2023evolutionary}, and social exclusion \cite{sasaki2013evolution,szolnoki2017alliance} based on evolutionary game theory. For instance, Xia \textit{et al.} reviewed the interplay of reputation and reciprocity in fostering cooperation, analyzing their dynamic definitions and effects on social dilemmas across different population structures while also synthesizing theoretical and experimental findings to outline promising future research directions \cite{xia2023reputation}. 

However, these traditional game models and mechanisms focus on pair interaction and ignore complex multi-person interactions such as family, teamwork, and community governance. Based on this, the spatial public goods game (SPGG) comes into being, which describes an interactive game between multiple individuals on the network. In such a multi-interaction, defectors will use the public pool to sacrifice the interests of cooperators like social welfare or environmental resources, leading to the ``tragedy of the commons'' \cite{hardin2018tragedy}. This dilemma presents significant challenges to the emergence of cooperation in SPGG, thereby stimulating extensive mechanism research on the subject \cite{wang2023inertia,ling2025supervised,zhang2025evolution}.
However, the traditional punishment mechanism often leads to the emergence of ``second-order free-riders'' who are willing to cooperate but unwilling to punish the defectors due to the high cost \cite{panchanathan2004indirect,fowler2005second}. Based on these findings, researchers have reevaluated the role of rewards in cooperation, showing that moderate rewards are more effective than high rewards, emphasizing the complexity of reward mechanisms and the impact of cost and synergy effects \cite{szolnoki2010reward}. In addition, Sun \textit{et al.} studied optimal incentive allocation in PGG on regular networks, deriving dynamical equations via pair approximation and formulating strategies to enhance cooperation by balancing rewards and punishments \cite{sun2023state}.

Among the various approaches to resolving social dilemmas, the stochastic process approach has gained attention due to its ability to capture the randomness in real-life phenomena, which is widely used in evolutionary game theory \cite{zeng2025complex,feng2024information}. In social problems such as public resource management, environmental governance, and the maintenance of cooperative behavior, individual behavior usually has uncertainty, which is often a direct manifestation of stochastic processes \cite{traulsen2009stochastic,tsur2021dynamic}. In evolutionary game theory, stochastic processes primarily model the updating of individual strategies, the adaptation of group structures, and the spread of cooperative behavior. Classical evolutionary dynamics models, such as replicator dynamics \cite{mendoza2024evolutionary,hofbauer2003evolutionary,li2023open}, Fermi update rule \cite{shen2024learning}, and best-response dynamics \cite{evilsizor2014evolutionary}, can be generalized in a stochastic environment to more realistically simulate the decision-making process of individuals under limited cognition and environmental noise. Drik developed a stochastic model of imitative interactions, deriving stochastic game dynamical equations and covariances, which influence average behavior and signal social phase transitions, with simulations demonstrating their role in behavioral self-organization \cite{helbing1996stochastic}. In particular, Feng \textit{et al.} proposed a game transition model based on the Markov process, which provides a comprehensive perspective for the emergence of cooperation \cite{feng2023evolutionary}.

In the classic SPGG, cooperation and defection are modeled as investment in or abstention from contributing to a public pool, with the benefits typically distributed instantaneously. That is, all players decide on their strategies simultaneously, and the public pool immediately allocates the resulting benefits. However, in real-world scenarios, player behavior is not synchronous. Instead, each individual's decision occurs at a specific point in continuous time. Focusing on this phenomenon, we propose a continuous SPGG based on a queueing system in this paper.  In summary, the main contributions of this paper are outlined as follows.

\begin{enumerate}
  \item Each player in the complex network will randomly enter the $M/M/1$ queueing system at a rate $\lambda$, and the service time of each player obeys the exponential distribution of $\mu$. The game will start after all players in the game group have completed the service, where the role of the service center is to determine the player's game status. 
    \item  The traditional fixed payoff function is modified by introducing a dependency between the enhancement factor and the sojourn time in the queueing system. Unlike conventional models where all cooperators receive the same enhancement, individuals who remain in the queue longer obtain greater enhancement, adding a dynamic aspect to the payoff structure.
    \item  The role of a reputation mechanism in fostering cooperation in the continuous SPGG is examined. Reputation accumulates through cooperation and decreases due to defection. A probability parameter is introduced to govern the selection mechanism, where, with a certain probability, an individual selects the player with the highest reputation, while with the complementary probability, a neighbor is chosen randomly. These modifications enhance the model’s realism by better capturing decision-making dynamics.
\end{enumerate}

The rest of this paper is structured as follows. Section \ref{model} presents the continuous SPGG framework based on queuing theory, covering the game model, queuing system, strategy update rules under the reputation mechanism, and theoretical analysis. Section \ref{simulation} provides the results of the experimental simulations along with a detailed analysis. Finally, this paper summarizes the key findings and explores possible avenues for future research in Section \ref{outlooks}.
\section{Model}\label{model}
In this section, we introduce an innovative continuous SPGG model. Building upon the traditional SPGG framework, we first develop an $M/M/1$ model to characterize the individual queueing service process, subsequently extending it to encompass the queueing dynamics of all individuals within the network. Following the game's completion, a reputation mechanism is employed to determine neighbor selection for payoff comparison, and the evolutionary process is then governed by the Fermi rule update strategy. Furthermore, we use the birth-death equation of individuals in the queueing system to construct a Markov chain and utilize the equilibrium equation to determine the system's stationary probabilities. This approach allows us to derive the total payoff pool within the square lattice network and systematically analyze the impact of queueing parameters on the level of cooperation. In the following subsection, we present the core components of the model.
\subsection{Spatial public goods game}
We first propose a standard SPGG within the framework of evolutionary game theory. The game is played on various types of complex networks, where nodes represent players and edges represent the interactions between them. Participants are classified into two groups: cooperators (\( s = C \)), who actively contribute to the public pool, and defectors (\( s = D \)), who refrain from contributing but still benefit from the distribution of the public pool. Within this framework, every node in the network serves as a focal point, establishing a game group alongside its interacting neighbors. All cooperators contribute a fixed amount \( c \), and the total contributions are amplified by an enhancement factor \( r \). The resulting pool is then equally distributed among all participants, regardless of their individual contributions. Consequently, the payoff of a player in the game of the focal node \( i \) is given by:
\begin{equation}
\left\{
\begin{aligned}
\Pi_{i}^C &= \frac{rc\left| \mathcal{N}_{i}^C \right|} {\left| \mathcal{N}_{i} \right|} - c ,\\
\Pi_{i}^D &= \frac{rc\left| \mathcal{N}_{i}^C \right|}{\left| \mathcal{N}_{i} \right|},
\end{aligned}
\right.
\label{Eq:1}
\end{equation}
where $\mathcal{N}_{i}^C$ denotes the set of cooperating individuals in the game group centered at node $i$, while $\mathcal{N}_{i}$ denotes the whole player set. Consequently, the total payoff for individual $i$, who participates in $\left| \mathcal{N}_{i} \right|$ group games, can be calculated as $\Pi_i= \sum_{j \in \mathcal{N}_{i}}\Pi_{j}^{s_i}$.

\subsection{SPGG with queueing system}
The traditional SPGG typically determines public pool allocation based on fixed inputs at discrete time intervals. To extend this framework to a continuous-time model, we introduce a queueing simulation that captures the contribution dynamics of cooperators and the free-riding behavior of defectors.

Before incorporating the queueing system into the SPGG, we first outline some necessary assumptions. Player inputs arrive at a rate of \( \lambda \), and these inputs are processed by a service center at a rate of \( \mu \) for payoff allocation.
Building upon this framework, we model each player as an independent customer, where the entire behavioral cycle, from game entry to departure, constitutes a queueing process. By integrating the principles of birth-death processes, we develop a mathematical representation of the system dynamics, characterizing the players' input and output patterns as a continuous-time Markov chain. The fundamental queueing framework is proposed as follows:

\begin{enumerate}
    \item \textit{Input process}: We utilize a homogeneous Poisson process to model player arrivals, where the parameter $\lambda$ represents the player arrival rate.

    \item \textit{Service process}: The system consists of an operational service center, which processes requests independently at a fixed service rate \( \mu \). As a result, the players' service times follow an exponential distribution, as described by \cref{E}.
\begin{equation}
P(t) = \left\{
\begin{array}{ll}
1 - \exp(-\mu t) & \quad t \geq 0, \\
0 & \quad \text{otherwise.}
\end{array}
\right.
\label{E}
\end{equation}

    \item \textit{Capacity}: The queueing system operates with a single finite queue. Given that the number of players in the network is finite (\( N \)), the total number of customer arrivals per round is inherently limited. Consequently, unlike traditional models that assume an infinite customer source, our framework explicitly accounts for this limitation. Nevertheless, the multi-round nature of the game allows players to reenter the system multiple times, ensuring that the queue dynamics are sustained despite the finite population.
    
    \item \textit{Queueing discipline}: Players enter the queueing system randomly and are assigned to an idle service center following a first-come-first-served (FCFS) discipline. 

\end{enumerate}

\begin{figure*}[h]
    \centering
    \includegraphics[width=1\linewidth]{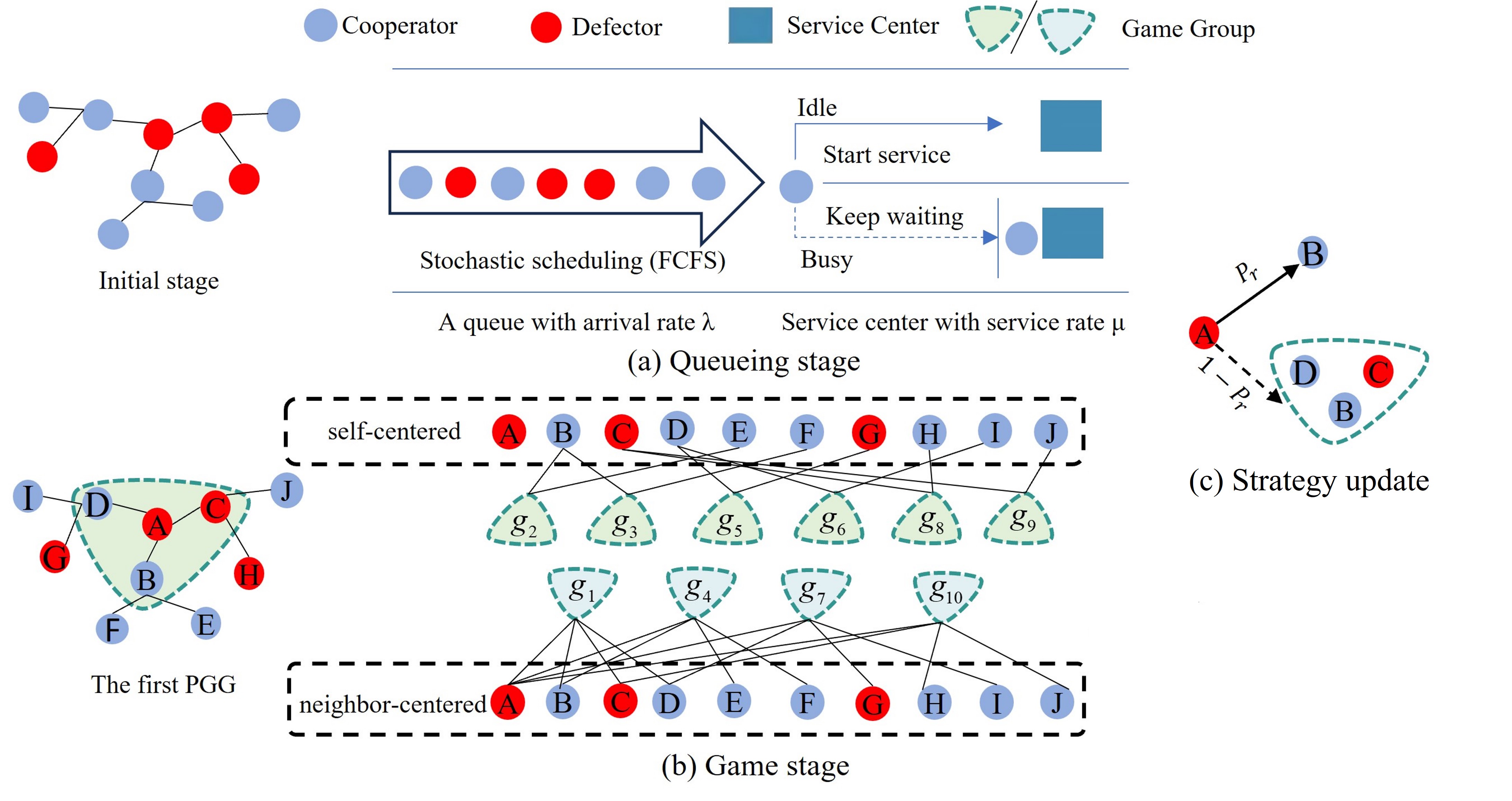}
    \caption{
\textbf{Schematic diagram of the proposed continuous SPGG model based on a queueing system.} The model consists of three main stages: (a) the queueing process of network players, (b) the sequence of public goods games, and (c) the strategy update process. Firstly, the initial strategies of players are randomly assigned. Secondly, players enter an \( M/M/1 \) queueing system with an arrival rate \( \lambda \) and are served with an exponential service rate \( \mu \). Upon completion of service, they participate in a PGG. Finally, in the strategy update stage, the focal node selects a neighbor with the highest reputation with probability \( P_r \) or randomly selects a neighbor with probability \( 1 - P_r \), then strategy update follows the Fermi function.
}
    \label{Fig: 1}
\end{figure*}
Following standard queueing theory notation, our proposed model is represented as an \( M/M/1/N/N \) system. In this notation, the first \( M \) indicates a Poisson arrival process for players, the second \( M \) signifies exponentially distributed service times, and \( 1 \) represents a single service center, which is a public institution used to determine whether an individual enters the game in continuous SPGG. The two \( N \) correspond to the system's queue capacity and the total customer population, respectively.


\subsection{Continuous SPGG with heterogeneous enhancement}
As shown in \cref{Eq:1}, the total enhancement in the game group is always a linear function of the number of cooperators and the enhancement factor \( r \). However, after incorporating the queueing system, the investment of cooperators depends on the sojourn time. In queueing theory, the sojourn time represents the total duration a customer remains in the system, encompassing both waiting time and service time. For the $M/M/1$ queueing system, the sojourn time $T_i$ for player $i$ is given by:
\begin{equation}
      T_i = W_i + S_i,
\end{equation}
where \( W_i \) denotes the queueing time of player \( i \), and \( S_i \) represents the corresponding service time. 

In real-world scenarios, although individuals may pay the same cost when participating in collective actions, substantial disparities often arise in the benefits they receive. These differences can be attributed to variations in resource acquisition, social status, ability levels, or external environmental factors.
It is assumed that each cooperator invests an amount \( c \) into the queueing system. However, the cooperator's payoff varies depending on their sojourn time in the system. This differential gain can be expressed as the product of the sojourn time and an enhancement factor, which is then distributed equally among all players. Consequently, the new payoff function for cooperator and defector can be defined as \cref{EE}.
\begin{equation}
\left\{
\begin{aligned}
\Pi_i^C &= \frac{rc  \sum_{j \in \mathcal{N}_{i}^{C}} T_j}{\left| \mathcal{N}_{i} \right|} - c ,\\
\Pi_i^D &= \frac{rc  \sum_{j \in \mathcal{N}_{i}^{C}} T_j}{\left| \mathcal{N}_{i} \right|}.
\end{aligned}
\right.
\label{EE}
\end{equation}

\subsection{Reputation assessment and strategy evolution}

In our continuous SPGG, we incorporate a reputation assessment mechanism for players. Each player's reputation is initially uniformly distributed within the range [0, 1] before the start of the iteration. After each round of the queueing process, players' reputations are updated based on their chosen strategies. The specific update rules are defined as follows:

\begin{equation}
    R_{i}(t)=
\begin{cases}
R_{i}(t - 1)+0.05 &\text{if } s_{i}(t)=C,\\
R_{i}(t - 1)/2 &\text{if } s_{i}(t)=D,
\end{cases} \label{2}
\end{equation}
where \( R_{i}(t) \) represents the reputation of individual \( i \) after \( t \) iterations. To simplify the model and prevent unbounded reputation growth, we constrain reputation values within the range \([0, 1]\).

We hereby adopt the classical Fermi update rule in continuous SPGG. First, the individual will select the player with the highest reputation by a certain probability \( P_r \) to compare the payoff; Otherwise, the individual will randomly select a neighbor. Then, the Fermi function given by \cref{Eq: 5} is used to transform the payoff difference into the update probability to determine whether the focal individual $i$ adopts the neighbor $j$'s strategy.
\begin{equation}
P ( s _ { i } \, \leftarrow \, s _ {j}) = \frac { 1 } { 1 + e ^ { ( \Pi _ { i } - \Pi_{ j } ) / \kappa } },
\label{Eq: 5}
\end{equation}
where the parameter $ \kappa$ represents the noise factor, which characterizes the uncertainty in strategy updates. When $ \kappa\rightarrow 0$, individual \( i \) will highly likely adopt the strategy of individual  \( j \) if \( j \) has a higher payoff. Conversely, when $ \kappa\rightarrow  \infty $, the strategy update becomes entirely random. Based on the existing evolutionary studies \cite{quan2020reputation,tang2024cooperative}, we usually set \(  \kappa  = 0.5 \).

To summarize our model, we present a schematic diagram of the continuous SPGG. Specifically, we illustrate the detailed process of individual queueing in the network in \cref{Fig: 1}(a). Individuals randomly enter the queueing system, which operates on an FCFS discipline, and exit after receiving service at the service center. In particular, when a player is being served at the service center, all players in the queue must wait. Once the service is completed, the leading player in the queue will proceed to receive service. Conversely, when the service center is idle, a player can directly enter and begin service.
Once all players in a game group have completed their respective service processes, the game begins.  

In the continuous SPGG, the sequence of games is determined by the queueing order of individuals. Upon completing their service, an individual first checks whether all members of the game group associated with their node have also completed their service. When this condition is met, a self-centered game is initiated, with the current node serving as the focal node of the game. Additionally, the individual evaluates whether their service completion enables any neighboring node to meet the conditions for initiating a game. In such cases, a neighbor-centered game is triggered, with the game being centered on the neighboring node rather than the current node. As illustrated in \cref{Fig: 1}(b), we suppose the queueing order of individuals is \( Q = (A, B, C, D, E, F, G, H, I, J) \). When players \( A, B, C \) complete the service, no game group satisfies the required conditions, and therefore no game occurs. However, when player \( D \) completes the service, the first neighbor-centered game \( g_1 = \{A, B, C, D\} \) is triggered by checking the game group centered on \( A \) for the first time. Among the queueing order \( Q \), \( (g_3, g_4) \), \( (g_6, g_7) \), and \( (g_9, g_{10}) \) are games triggered by nodes \( F \), \( I \), and \( J \), respectively. This indicates that the service of these nodes not only initiates a self-centered game but also satisfies the conditions for a neighbor-centered game. In all games, \( \{ g_2, g_3, g_5, g_6, g_8, g_9\} \) correspond to self-centered games, and \( \{ g_1, g_4, g_7, g_{10} \} \) represent neighbor-centered games. 

After all players in the network have entered and exited the $M/M/1$ queuing system, players will select neighbors to compare payoff as shown in \cref{Fig: 1}(c). Specifically, individual A will choose the player $B$ with the highest reputation among its neighbors with probability \( P_r \) to compare payoff. Otherwise, a player is randomly selected from its neighboring group \( \{ B, C, D \} \) with probability \( 1 - P_r \). 
Similarly, all players in the network will select neighbors to compare payoff in the same way, and ultimately, the Fermi rule will be applied to decide whether to update their strategy.

\subsection{A brief analysis of continuous SPGG based on $M/M/1$ queueing system}
To further analyze the stochastic dynamics of the continuous SPGG with a queueing system, we develop a birth-death process framework to derive the stationary probabilities for various system states. In this queueing-based SPGG, players arrive at the system at a rate \( \lambda \) and are serviced at a rate \( \mu \). Each player can be modeled as a customer entering a queueing system, and the overall process is represented as a continuous Markov chain. According to the balance equations of the birth-death process, we have

\begin{equation}
    \begin{cases}
        \mu P_1 = \lambda P_0 & n = 0, \\
        \lambda P_{n-1} + \mu P_{n+1} = (\lambda + \mu) P_n & 1 \leq n \leq N-1,\\
         \lambda P_{N-1} = \mu P_N & n = N,
    \end{cases}
\label{Eq: 2}
\end{equation}
where \( P_{n} \) denotes the probability of the system being in state \( n \) at equilibrium. The objective of this study is to examine the scenario where the queue length in the queueing system is finite, which implies that the player can transition between service states within a restricted time frame. In this context, the system's load, denoted by \( \rho = \frac{\lambda}{\mu} \), is constrained to be less than 1. By applying the normalization condition \( \sum_{n=0}^{N} P_{n} = 1 \), the stationary probability of having \( n \) individuals in the queueing system is given by

\begin{equation}
P_n = \frac{(1 - \rho) \rho^n}{1 - \rho^{N+1}} \quad 0 \leq n \leq N.
  \label{Eq: 8}
\end{equation}

Our purpose is to explore the number of players, so we can use the queueing system indicators to reflect the game's attractiveness to the group. Based on the stationary distribution obtained by \cref{Eq: 8}, we can obtain the average queue length of the system by

\begin{equation}
\begin{aligned}
L &= \sum_{n=0}^{N} n P_n = \frac{\rho}{1 - \rho} - \frac{(N+1) \rho^{N+1}}{1 - \rho^{N+1}}.
\end{aligned}
\end{equation}

Similarly, the average sojourn time of the customer can also reflect the crowding degree of the game and the trend of the player's sojourn time. According to Little's equation, the average sojourn time of the player can be calculated as follows:
\begin{equation}
\begin{aligned}
E(T) &= \frac{L}{\lambda} = \frac{1}{\lambda} \left( \frac{\rho}{1 - \rho} - \frac{(N+1) \rho^{N+1}}{1 - \rho^{N+1}} \right).
\end{aligned}
\end{equation}

\begin{algorithm}[t]
    \caption{Evolutionary Game Simulation with Queueing System}
    \label{alg:EvolutionaryGame}
    \renewcommand{\algorithmicrequire}{\textbf{Input:}}
    \renewcommand{\algorithmicensure}{\textbf{Output:}}
    \begin{algorithmic}[1]
        \REQUIRE Network $G$, time step $t$, enhancement factor $r$, arrival rate $\lambda$, service rate $\mu$  
        \ENSURE cooperation proportion $\rho_{c}$ 
        
        \STATE Assign random strategies to nodes in $G$   
        \STATE Set payoff of each node to 0  
        
        \FOR{each Monte Carlo time step $t$}  
            \STATE Generate arrival times and service times using exponential distributions with arrival rate $\lambda$ and service rate $\mu$  
            \STATE Calculate service start and end times  
            \STATE Mark nodes as ``completed'' when service ends  
            
            \FOR{each node and its neighbors}  
                \IF{all nodes in the game group complete service}  
                    \STATE Calculate player's payoff and reputation 
                \ENDIF
            \ENDFOR
            
            \STATE Calculate cooperation proportion $\rho_{c}$  
            \IF{Cooperation proportion $\rho_{c}$ is 0 or 1}  
                \RETURN cooperation proportion $\rho_{c}$  
            \ENDIF
            
            \FOR{each selected node in the  network $G$}  
                \STATE Compare payoff with the selected neighbor  
                \STATE Use Fermi function to decide strategy updates  
            \ENDFOR
            
        \ENDFOR
        
        \RETURN Average cooperation proportion $\rho_{c}$ of the last 500 iterations  
    \end{algorithmic}
\end{algorithm}

Traditional research on square lattice networks has proved that the cooperation level of the network depends on the enhancement factor $r$. We apply the continuous SPGG to square lattices with periodic boundary conditions, where each player interacts with its four nearest neighbors defined by the von Neumann neighborhood. Based on the framework, the total enhancement of all public pools in the network can be expressed by

\begin{equation}
\begin{aligned}
\psi  &= \sum_{i=1}^N \sum_{j \in \mathcal{N}_i^C} r\cdot T_j \\
&= 5r\cdot N_{c}E(T) \\
&= 5rN_{c}\frac{1}{\lambda} 
\left( \frac{\rho}{1 - \rho} - \frac{(N+1) \rho^{N+1}}{1 - \rho^{N+1}} \right),
\end{aligned}
\end{equation}
where $\psi $ denotes the total enhancement of the game in the network and  $N_{c}$ represents the number of cooperative players in the whole network. 

To further explore the impact of queueing dynamics on the system, we consider the limiting case where the network size \( N \) becomes large. Taking the limit \( N\to\infty \) in the expression for \( \psi  \), we observe that for \( \rho < 1 \), the total enhancement of the square lattices network is simplified to

\begin{equation}
\lim_{N \to \infty} \psi  = \lim_{N \to \infty}5r\cdot N_{c}E(T)=\frac{5rN_{c}}{\mu-\lambda}.\label{bb}
\end{equation}

Compared to the traditional SPGG, our model integrates queueing parameters to investigate their influence on the public pool’s gain in continuous time. Specifically, a queueing system with a high arrival rate and a low service rate extends players' sojourn time in the queue, thereby enhancing their cumulative contributions to the game. As a result, the total revenue of the network increases, and the payoff gap between cooperators and defectors is reduced, thereby promoting the emergence of cooperation.
\begin{figure*}[h]
    \centering
    \includegraphics[width=1\linewidth]{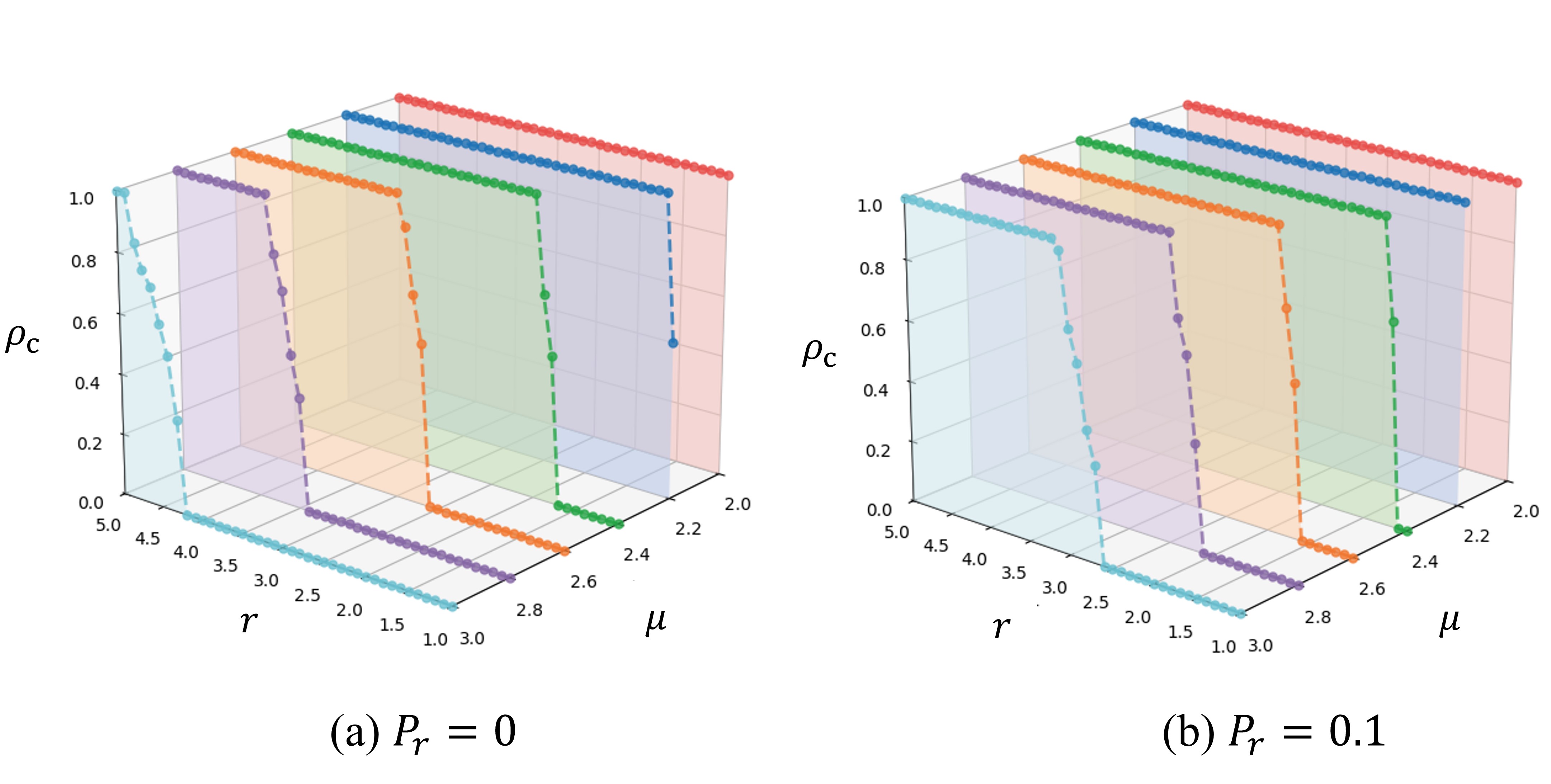}
    \caption{\textbf{The curve of cooperation ratio \( \rho_{c} \) with \( r \) under different service rates \( \mu \).} The arrival rate is fixed at \( \lambda = 2 \). Different panels illustrate the cooperation level for varying probabilities \( P_{r} \) of selecting the highest-reputation player: (a) \( P_{r} = 0 \), (b) \( P_{r} = 0.1 \). These curves depict the transition dynamics from cooperation to defection within the network under different service rates.}
    \label{Fig: 2}
\end{figure*}
\section{Simulation results and discussions}
\label{simulation}
In this section, we conduct simulations to validate our model. Specifically, we first outline the simulation methodology. Based on this framework, we then investigate the influence of the enhancement factor and queueing parameters on the evolutionary trajectory and the density of cooperation within the network.

\subsection{Experimental method and parameter setting}
The Monte Carlo simulations in this paper are primarily conducted on square lattices with periodic boundary conditions, where the network size is set to \( N = 50 \times 50 \). Initially, each individual's strategy is randomly assigned, and their reputation values are uniformly distributed within the range of \( [0,1] \).  After the strategy update of each Monte Carlo time step $t$, the node's state reverts to unserviced and reenters the queueing system. At the same time, the queueing time and service time of the individual are reset to 0.

Our research focuses on the level of cooperation in the network. By integrating the strategy update mechanism with the established continuous SPGG, our evolutionary computation algorithm for stable cooperation proportions can be represented as Algorithm 1. For the termination time of the SPGG, we typically set a sufficiently large \( t \) to ensure that the network evolves into one of three possible steady states: pure cooperation, pure defection, or a stable coexistence of both strategies. To ensure the accuracy of the final results of the evolutionary process, we average the outcomes over 10 independent experiments.

\subsection{Influence of enhancement factor and reputation on
cooperation}

In this subsection, we examine how the cooperation level is influenced by the enhancement factor \( r \) under the reputation mechanism. As shown in \cref{Fig: 2}, regardless of whether the reputation mechanism is present, the overall cooperation level in the network increases as the enhancement factor \( r \) grows. When \( P_r = 0 \) in \cref{Fig: 2}(a), the population eventually consists entirely of defectors for small values of \( r \), indicating a state of pure defection when \( \mu = 2.4, 2.6, 2.8\) and \( 3.0 \). As \( r \) increases, a transition occurs where cooperators and defectors begin to coexist within the network. Specifically, for the four curves corresponding to \( \mu = 2.4, 2.6, 2.8, \) and \( 3.0 \), the critical values of \( r \) at which the system shifts from pure defection to a mixed state are \( r = 1.7, 2.6, 3.4, \) and \( 4.2 \), respectively. This indicates that as \( \mu \) increases, a higher enhancement factor \( r \) is required to promote the emergence of cooperation. In particular, when \( \mu = 2.2 \), the network exhibits two distinct states: coexistence of cooperators and defectors, and pure cooperation. When \( \mu = 2 \), the network reaches a state of pure cooperation.
\begin{figure*}
    \centering
    \includegraphics[width=1\linewidth]{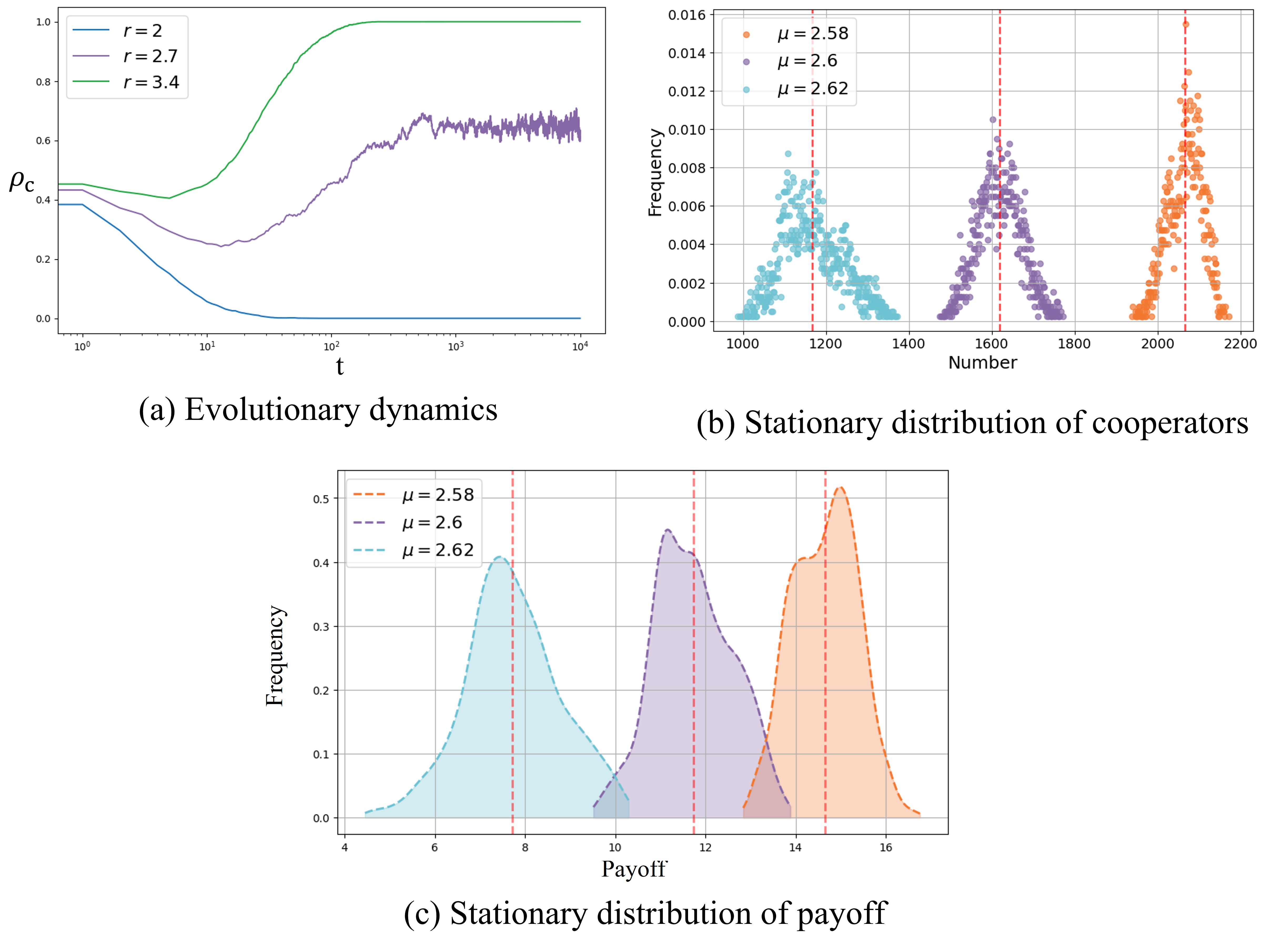}
    \caption{\textbf{Evolutionary dynamics, stationary distribution, and payoff distribution of players. }The arrival rate is fixed at \( \lambda = 2 \) and $P_r=0$.
    (a) The evolution of cooperation ratio $\rho_c$ over time $t$ for different values of $r$. 
    (b) The stationary distribution of cooperative players for $r=2.7$ under different values of $\mu$.  The data point represents the distribution of the number of cooperators over the last 4000 steps. 
    (c) The probability density distribution of payoff in the final stage for $r=2.7$ . The red dashed lines in (b) and (c) represent the mean values of the number of cooperators and the payoff under the corresponding parameters, respectively.}
    \label{Fig: 3}
\end{figure*}

When the reputation mechanism is introduced, as shown in \cref{Fig: 2}(b), the critical values of \( r \) at which the network transitions from a state of pure defection to the strategy coexistence state for \( \mu = 2.4, 2.6, 2.8 \), and \( 3.0 \) are \( r = 1.1, 1.6, 2.1 \), and \( 2.6 \), respectively. This demonstrates that the presence of a reputation mechanism lowers the threshold of \( r \) required for the emergence of cooperation, thereby facilitating the spread of cooperative behavior in the network. Similarly, for the queueing system with a lower service rate, where the sojourn time results in higher payoffs, the network only consists of pure cooperators when \( \mu = 2 \) and \( \ 2.2 \).

By studying the influence of \(r\) and \(\mu\) on the level of group cooperation with and without the reputation mechanism, we conclude that both the enhancement factor \(r\) and the service rate \(\mu\) play a pivotal role in shaping the evolutionary trajectory of cooperation. The introduction of a reputation mechanism significantly lowers the critical values of \(r\) required for the emergence and dominance of cooperation, thereby promoting cooperative behavior more effectively. Furthermore, our findings indicate that for smaller values of \(\mu\), cooperation is more easily sustained, and the system quickly transitions from a state of defection to coexistence and eventually to full cooperation. In contrast, for larger \(\mu\), a higher enhancement factor \(r\) is required to achieve the same level of cooperation. This suggests that the synergy effect in collective interactions plays a crucial role in determining the stability and prevalence of cooperative strategies within the network.

\subsection{Evolution of strategy on the square lattice}
In evolutionary game theory, analyzing the change in the cooperation level \(\rho_c\) over time is of great significance for understanding the dynamic process of strategy evolution. By examining the trajectory of \(\rho_c\), we can assess the stability and persistence of cooperation under different conditions, as well as the impact of key parameters on the long-term behavior of the system. This analysis enables us to identify the conditions under which cooperation becomes sustainable, thus providing deeper insights into the emergence of cooperation within the network.

\begin{figure*}[h]
    \centering
    \includegraphics[width=1\linewidth]{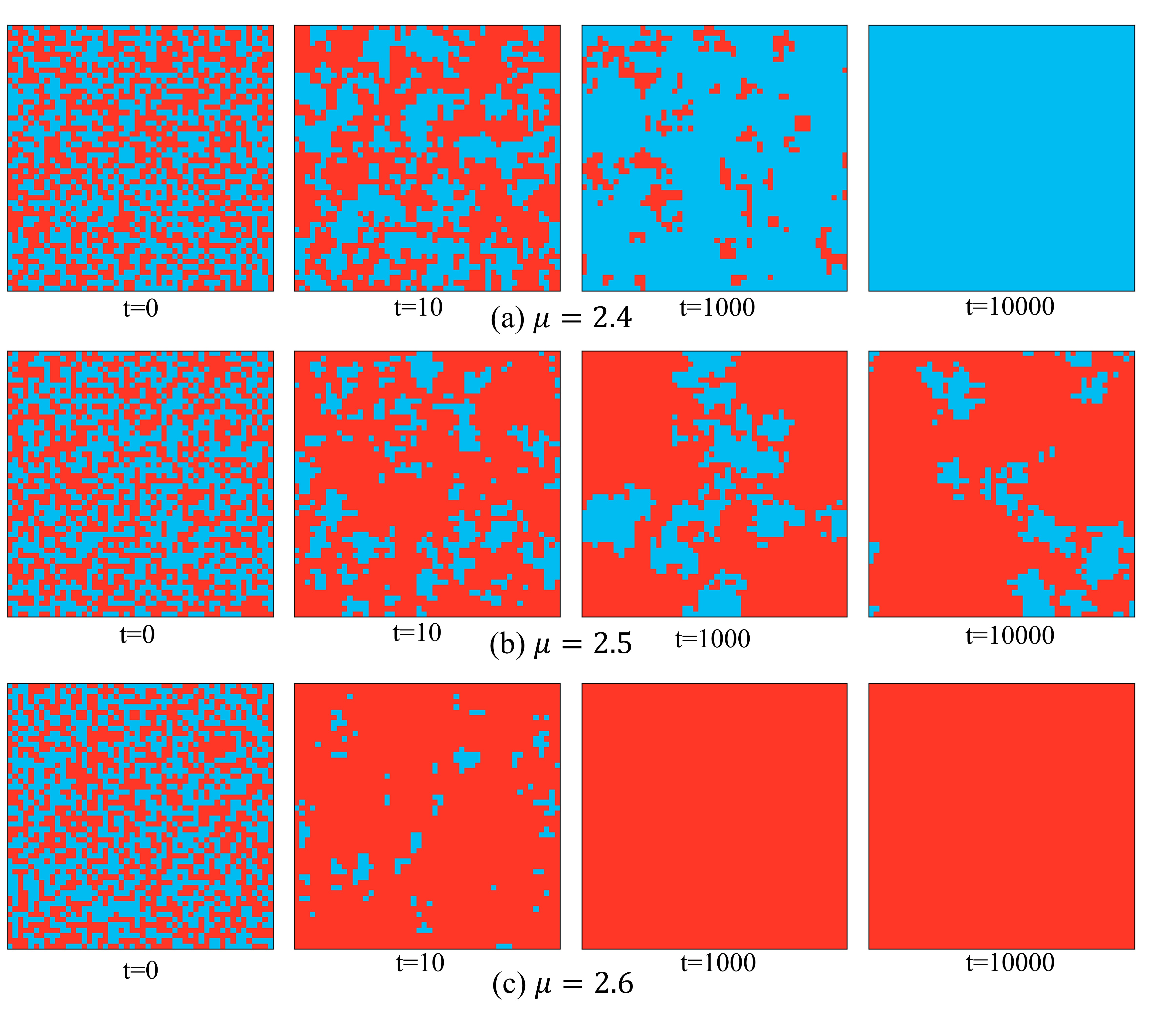}
    \caption{\textbf{Evolutionary snapshots under different service rates \(\mu\)}. Each row from top to bottom represents a situation corresponding to different service rates \(\mu\). Each column from left to right represents different time steps \(t\). Blue pixels represent cooperators, while red pixels represent defectors. The parameters are fixed with \(r=2.2\), \(P_r=0\), and \(\lambda=2\). (a) \(\mu=2.4\); (b) \(\mu=2.5\); (c) \(\mu=2.6\). }
    \label{Fig: 4}
\end{figure*}

In \cref{Fig: 3}(a), we show the change curve of the proportion of cooperators with the strategy update step $t$ under different enhancement factors $r$. When \(r = 2.7\) and \(r = 3.4\), the proportion of cooperators initially decreases to its lowest point between \(t = 10\) and \(t = 100\), then gradually increases after \(t = 100\) and stabilizes by \(t = 1000\). The population exhibits no defection strategy for \(r = 3.4\) when the evolution reaches a stable state. However, for \(r = 2.7\), a coexistence of cooperators and defectors is observed, as indicated by the fluctuating cooperation ratio. This suggests intense competition between cooperators and defectors within the network. Notably, when \(r\) is relatively low, the proportion of cooperators declines sharply to zero, implying that a low enhancement factor \(r\) favors the survival of defectors. To further investigate the impact of the queueing system on the coexistence state, we vary the service rate \(\mu\) in the coexistence state depicted in \cref{Fig: 3}(a). The results align with \cref{Fig: 2}, indicating that a lower service rate \(\mu\) fosters the emergence of cooperation. This phenomenon can be attributed to increased service center congestion and longer sojourn time for cooperators within the system. As the service rate \(\mu\) decreases, the time spent by cooperators in the system increases, which results in higher payoff due to the extended participation in the game. This increased payoff enhances the incentives for cooperation, facilitating the emergence and stabilization of cooperative behavior in the network. 

Interestingly, as illustrated in \cref{Fig: 3}(b), we observe that lower values of \(\mu\) have a smaller impact on the volatility of the final cooperator population, which is represented by the number of cooperators being more tightly clustered around the mean. It can be seen from \cref{Fig: 3}(c) that a high service rate leads to an increase in the variance of payoff. Since the Fermi update rule we employ is based on payoff differences, higher service rates facilitate more frequent and easier strategy switching between cooperators and defectors. This increased strategy switching intensifies the competition between cooperators and defectors when the system reaches a stable state. Consequently, the fluctuation in the proportion of cooperators becomes more pronounced, resulting in greater volatility in the final number of cooperators. This suggests that a higher service rate can destabilize cooperation and lead to a more dynamic, competitive environment.

\begin{figure*}[!h]
    \centering
    \includegraphics[width=1\linewidth]{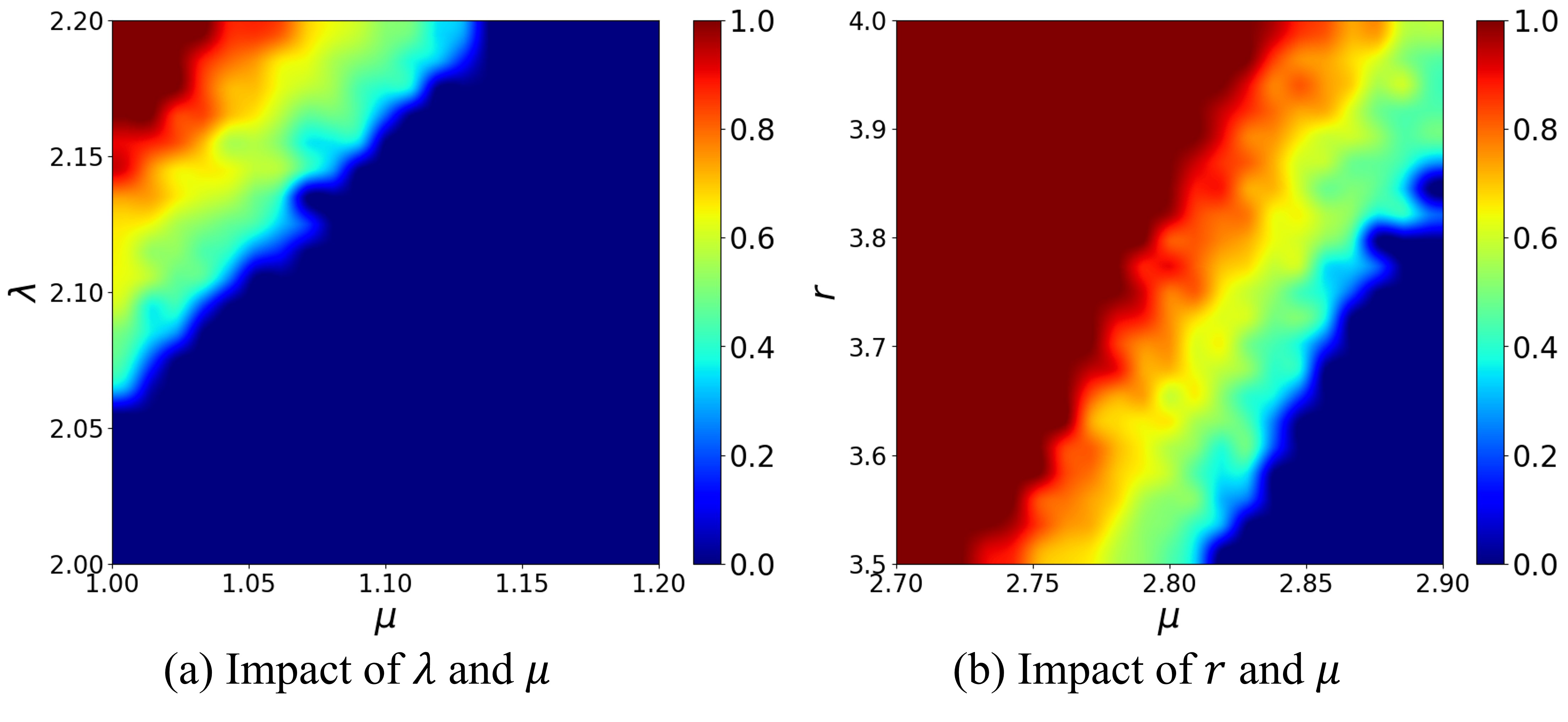}
    \caption{\textbf{The heatmap of cooperation level under different parameter settings on square lattice network}. The density of cooperators $\rho_{c}$ is shown on different parameter planes. The illustration on the right side of the panel explains the meaning of color. (a) The impact of service rate $\mu$ and arrival rate $\lambda$ for fixed $r=4$ and $P_r=0$. (b) The influence of enhancement factor $r$ and service rate $\mu$ for fixed $\lambda=2$ and $P_r=0$.}
    \label{Fig: 5}
\end{figure*}

\subsection{Spatial snapshots of strategy evolution}
To gain deeper insights into the rise of cooperation and the development of cooperative clusters within the population, \cref{Fig: 4} shows the spatial distribution of defectors and cooperators on the square lattice network. By examining the strategy distribution at discrete time steps under different service rates \(\mu\), as shown in \cref{Fig: 4}, we can gain deeper insights into the dynamic process of strategy formation. These spatial distributions reveal how cooperation evolves over time and how the service rate influences the formation of cooperative and defective regions within the network.
At the initial time \( t = 0 \), the distribution of cooperators and defectors is random, with no noticeable clustering of either strategy. This randomness indicates that, at the outset, neither cooperators nor defectors possess a strategic advantage, and the population exists in a mixed state. 
As the simulation progresses to \( t = 10 \), small clusters of cooperators begin to emerge, likely as cooperators recognize and associate with one another, promoting cooperative behavior. However, these clusters remain sparse and do not dominate the network. In contrast, defectors continue to be widespread, exploiting cooperators to maximize their payoffs.
This exploitation is facilitated by the average distribution rule of the SPGG, resulting in the formation of defector clusters that encircle the cooperator clusters.
When the service rate \( \mu \) is set to 2.4, a notable phenomenon occurs, wherein cooperators gradually encircle defectors, ultimately leading to the complete dominance of cooperators across the network by \( t = 10000 \), as shown in \cref{Fig: 4}(a).
In contrast, when \( \mu = 2.5 \), a distinct outcome arises, where cooperators are encircled by defectors at \( t = 10000 \) in \cref{Fig: 4}(b), leading to a stable coexistence of both strategies within the network.
In particular, the defection strategy will quickly occupy the entire network when it evolves to $t = 1000$ when $\mu=2.6$ in \cref{Fig: 4}(c), which demonstrates that a high service rate $\mu$ diminishes the advantages of cooperators within the population, thereby facilitating the dominance of defectors across the entire network.

\begin{figure*}[!h]
    \centering
    \includegraphics[width=1\linewidth]{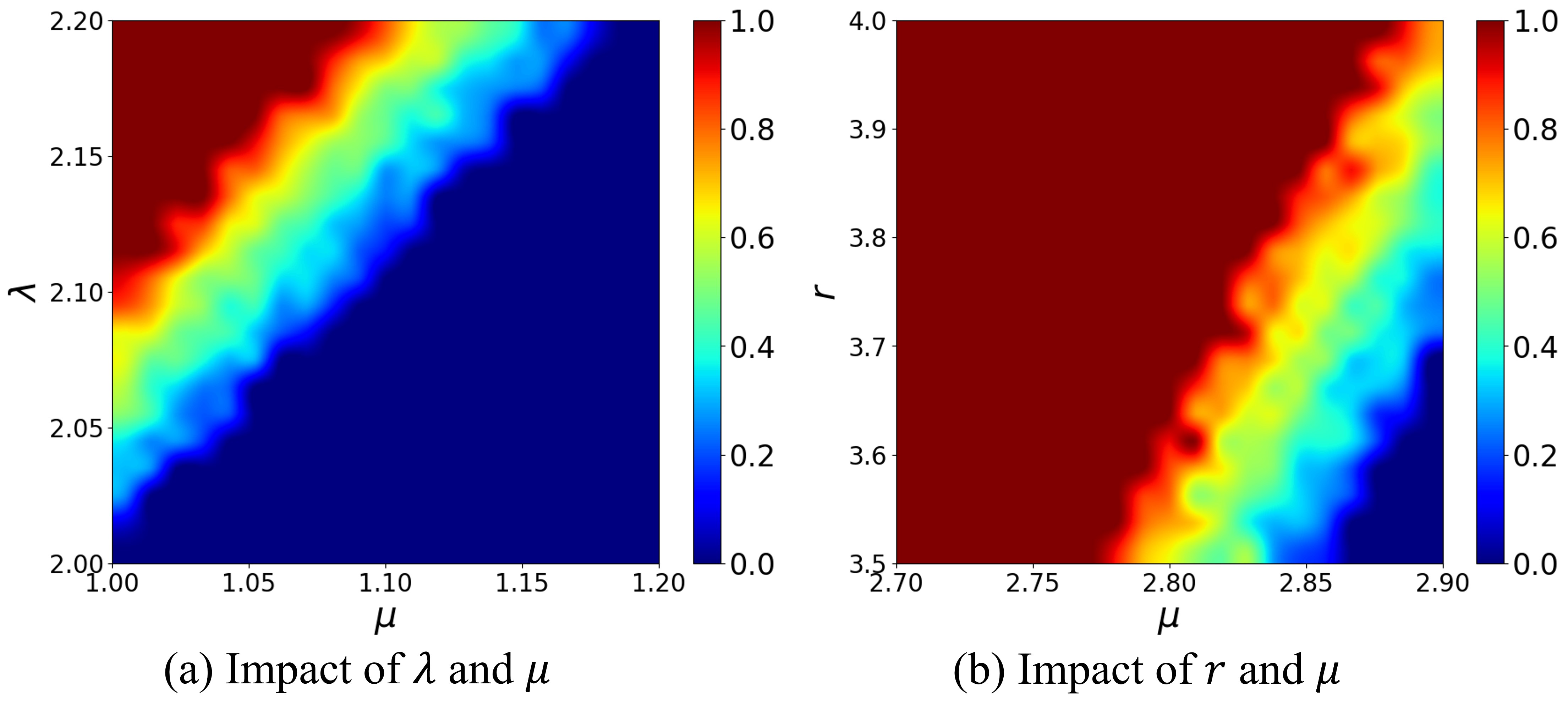}
    \caption{
\textbf{The heatmap of cooperation level under different parameter settings on small-world network}. The density of cooperators $\rho_{c}$ is shown on different parameter planes. The illustration on the right side of the panel explains the meaning of color. (a) The impact of service rate $\mu$ and arrival rate $\lambda$ for fixed $r=4$ and $P_r=0$. (b) The influence of enhancement factor $r$ and service rate $\mu$ for fixed $\lambda=2$ and $P_r=0$. The small-world network is generated using the Watts-Strogatz model with $N$ nodes, where each node initially has $k=4$ nearest neighbors and the rewiring probability $p=0.2$.}
    \label{Fig: 6}
\end{figure*}
\subsection{The comprehensive influence of queueing system on cooperation}
In our continuous SPGG model, previous simulations have shown that the parameters of the queueing system and the enhancement factor have a significant impact on the level of cooperation. Therefore, this subsection presents the cooperation levels under different parameter combinations and explores their interactions.

In \cref{Fig: 5}, we display the cooperative heatmaps for the square lattice network under varying service rate $\mu$ with rate $\lambda$, and enhancement factor $r$. \cref{Fig: 5}(a) reveals a distinct transition from a high level of cooperation to a low level along specific boundaries.
As \(\mu\) increases, cooperation tends to decrease, indicating that a higher service rate introduces greater competitive pressure on the system, thereby diminishing cooperative behavior. In contrast, as \(\lambda\) increases, cooperation becomes more prevalent, suggesting that a higher influx of players into the queueing system promotes cooperative interactions.
Furthermore, when the difference between \( \mu \) and \( \lambda \) remains constant, the level of cooperation remains virtually unchanged, consistent with the theoretical analysis presented in \cref{bb}.
In \cref{Fig: 5}(b), the results reveal additional states within the coexistence region. When the arrival rate \(\mu\) is low, individuals in the network tend to adopt the cooperation strategy due to its high returns. As \(\mu\) reaches a certain threshold, an increase in the enhancement factor \(r\) promotes the emergence of cooperation, leading to a mixed state of both strategies. However, when \(\mu\) exceeds a critical threshold, defectors dominate the entire network.

To further validate the rationality of our continuous SPGG, we have examined the impact of relevant queuing parameters on the level of network cooperation in a small-world network in \cref{Fig: 6}, which more closely resembles a real-world network. 
In this small-world network, each node has \( k=4 \) nearest neighbors, and the network's rewiring probability is \( p=0.2 \).
The simulation results are similar to those observed in the square lattice network. However, the cooperative region in the small-world network is significantly larger, indicating that the small-world topology facilitates a broader range of cooperative behavior under varying queuing parameters. In general, the addition of long-range edges in the small-world network allows defection groups to adopt cooperative strategies through these edges, which enhances the spread of cooperation strategies across the network \cite{xiong2024coevolution}.

This subsection examines the impact of queuing parameters on the level of cooperation in square lattice network and small-world network. The simulation underscores the complex interplay between competitive pressure and cooperative incentives within the system, offering a comprehensive   perspective on the emergence of cooperation in continuous SPGG.
\section{Discussion}\label{outlooks}
In this study, we propose a continuous SPGG framework that incorporates queuing systems to model dynamic and asynchronous interactions between individuals. By integrating the \(M/M/1\) queuing system with a reputation-based selection mechanism, our model effectively captures the stochastic nature of public goods provision in real-world social and economic systems. 
Extensive Monte Carlo simulations reveal that queuing parameters play a significant role in influencing the level of cooperation. The findings indicate that a high enhancement factor, elevated arrival rate, low service rate, and the implementation of a reputation mechanism collectively foster the emergence and stability of cooperative behavior within spatial networks.
Furthermore, our results offer a novel perspective on the interplay between time-dependent interactions and cooperation dynamics, emphasizing the critical role of individuals' sojourn time within the system and the influence of individual reputation in sustaining cooperation.

Future research could extend our model in several directions. For instance, incorporating temporal networks which consider that the structure of the network changes over time \cite{li2017fundamental}, may provide deeper insights into the role of topological features in facilitating cooperation \cite{li2020evolution}.
Furthermore, investigating the interplay of mechanisms like punishment, reward, and environmental feedback within our framework could further advance the understanding of cooperative behavior in dynamic and uncertain environments. Overall, the continuous SPGG can be extended to model more complex real-world scenarios by incorporating multi-interaction models, using stochastic approach for theoretical analysis, and further exploring the mechanisms underlying the emergence of cooperation.
\section*{ACKNOWLEDGMENTS}

G. Zhang, X. Xiong, B. Pi and M. Feng are supported by the National Natural Science Foundation of China (grant no. 62206230) and the Natural Science Foundation of Chongqing (grant no. CSTB2023NSCQ-MSX0064). M. Perc is supported by the Slovenian Research and Innovation Agency (Javna agencija za znanstvenoraziskovalno in inovacijsko dejavnost Republike Slovenije) (grant no. P1-0403).
\bibliographystyle{unsrtnat}


\bibliography{cas-refs}

\end{document}